\documentclass[12pt]{article}
\begin{document}
\def\a{\alpha}
\def\b{\beta}
\def\c{\varepsilon}
\def\d{\delta}
\def\e{\epsilon}
\def\f{\phi}
\def\g{\gamma}
\def\h{\theta}
\def\k{\kappa}
\def\l{\lambda}
\def\m{\mu}
\def\n{\nu}
\def\p{\psi}
\def\q{\partial}
\def\r{\rho}
\def\s{\sigma}
\def\t{\tau}
\def\u{\upsilon}
\def\v{\varphi}
\def\w{\omega}
\def\x{\xi}
\def\y{\eta}
\def\z{\zeta}
\def\D{{\mit \Delta}}
\def\G{\Gamma}
\def\H{\Theta}
\def\L{\Lambda}
\def\F{\Phi}
\def\P{\Psi}
\def\S{\Sigma}

\def\o{\over}
\def\beq{\begin{eqnarray}}
\def\eeq{\end{eqnarray}}
\newcommand{\gsim}{ \mathop{}_{\textstyle \sim}^{\textstyle >} }
\newcommand{\lsim}{ \mathop{}_{\textstyle \sim}^{\textstyle <} }
\baselineskip 0.7cm

\begin{titlepage}

\begin{flushright}

\end{flushright}

\vskip 1.35cm
\begin{center}
{\large \bf
Marginally Eternal Inflation
}
\vskip 1.2cm
Izawa~K.-I.
\vskip 0.4cm

{\it Yukawa Institute for Theoretical Physics, Kyoto University,\\
     Kyoto 606-8502, Japan}

\vskip 1.5cm

\abstract{
We examine the amount of parameter tuning
to make slow-roll inflation marginally eternal
by considering an $R$-invariant chaotic inflation
as an example of one-parameter tuning in supergravity.
The primordial inflation turns out to be possibly marginal
in such a setup.
}
\end{center}
\end{titlepage}

\setcounter{page}{2}


Realization of slow-roll inflation
\cite{Lyt,Gut}
implies parameter tuning of an inflaton potential
in order to make it sufficiently flat.
It seems naively that such tuning knows no bounds
\cite{Vil},
since the flatter the potential is, the longer the inflation lasts.
How flat is the inflaton potential?
Is it almost completely flat?

The minimal requirement of inflationary selection
\cite{Iza,Yan}
is that the potential is so flat as
to induce inflation indeed. Observationally,
the potential of the primordial inflation
should be flat enough to realize several tens of $e$-folds.

In this note, we examine the amount of parameter tuning
to make slow-roll inflation marginally eternal.
Namely, we consider marginal inflation, where
less tuned potential only realizes non-eternal inflation,
or irrelevant inflation,
and more tuning leads to relevant inflation.%
\footnote{The nomenclature is borrowed from that in
renormalization theory such as (non-)renormalizable interactions,
or (ir)relevant ones.}

Let us first present an $R$-invariant inflation model%
\footnote{See
Ref.\cite{Kaw,Yam}.
$R$-invariance is also utilized in other models
\cite{Iza,Izaw}.
}
of the chaotic type
as a concrete example of one-parameter tuning
in supergravity, which may be of some interest in its own right.
The superpotential $W$ and the K{\" a}hler potential $K$
of the model is given by
\beq
 W={\tilde m}X{\tilde \f} {\cal F}({\tilde \f}^2), \quad
 K=|X|^2 + 2A^2{\tilde x}^2 + 2{\tilde y}^2 + \cdots,
\eeq
where ${\tilde m}$ and $A$ are positive constants,%
\footnote{
The parameter $A$ is the one to be tuned to realize inflation
\cite{Shi}.
}
$X$ and ${\tilde \f}$ denote chiral superfields
with ${\tilde \f}={\tilde x}+i{\tilde y}$,
and ${\cal F}({\tilde \f}^2)=1+{\cal O}({\tilde \f}^2)$
is a generic holomorphic function of ${\tilde \f}^2$
in the reduced Planck unit.
The $R$-charges of ${\tilde m}X$ and ${\tilde \f}$
are two and vanishing, respectively,
with their parities odd for a $Z_2$ symmetry.
The ellipsis denotes generic higher-order terms respecting the symmetries.

For the normalized superfield%
\footnote{The variation of $\f$ is as large as $A > 1$
for that of a fundamental variable $|{\tilde \f}|<1$. 
}
$\f = \sqrt{A^2+1}{\tilde \f}$
with $\f = x+iy$,
we obtain
\beq
 W=mX\f {\cal F}\left( {\f^2 \o A^2+1} \right), \quad
 K=|X|^2 + {2A^2 \o A^2+1}x^2 + {2 \o A^2+1}y^2 + \cdots,
\eeq
where $m = {\tilde m}/\sqrt{A^2+1}$
and the ellipsis denotes higher-order terms
with suppressed variables $x/\sqrt{A^2+1}$ and $y/\sqrt{A^2+1}$
for large $A$.
The kinetic terms are approximately canonical
for small $|X|$ and $|\f|/A$.

The potential in supergravity for $\v=\sqrt{2}y$
under $x=X=0$ is given by
\beq
 V={1 \o 2} m^2\v^2 \left\{
   \left| {\cal F}\left( {\v^2 \o 2(A^2+1)} \right) \right|^2
   \exp \left( {\v^2 \o A^2+1} \right) + \cdots \right\}
  ={1 \o 2}m^2\v^2 \left\{ 1+{\cal O}\left( {\v^2 \o A^2+1} \right) \right\},
 \label{POT}
\eeq
which induces chaotic inflation
for $|\v| \lsim A$ with large $A \gg 1$.
Note that the approximately%
\footnote{If we could set up the superpotential linear in $\f$
and the K{\" a}hler potential independent of $y$,
we would obtain the quadratic potential as in
Ref.\cite{Yam}.
}
quadratic form of the inflaton potential
is predicted in this framework provided $A$ is sufficiently large.

We now present the condition that such a quadratic potential
supports eternal phase
\cite{Gut}
at $\v=aA \gg 1$ for a positive constant $a$.
The quantum fluctuations are given by
\beq
 \D \v \simeq {m \o 2\pi \sqrt{6}}aA. 
\eeq
On the other hand, the slow-roll during the Hubble time
is given by
\beq
 \d \v \simeq {2 \o aA}.
\eeq
The slow-roll tends to be compensated by the
quantum fluctuations for
$\D \v \gsim \d \v$,
namely,
\beq
 (aA)^2 \gsim 4\pi \sqrt{6} m^{-1}.
\eeq
The marginal inflation is realized when this condition is
marginally satisfied for the upper bound $a$ of order one
to induce inflation.%
\footnote{This condition might be regarded to imply
large $m$ instead of large $A$. However, high-energy-scale inflation
may be primary
\cite{Lyt,Izawa}
rather than primordial
under generic multiple inflations
\cite{Izawa,Kawa}
in the universe.
We here think of ${\tilde m}$ as given and $A$ to be tuned.}

For the case of primordial chaotic inflation
with $m \sim 10^{-5}$, the marginal parameter is given by
$aA \sim 10^3$, which implies a well-suppressed $\v^4$ term
in the potential Eq.(\ref{POT})
for several tens of $e$-folds.
If the primordial inflation is marginal,
such deviation from the quadratic potential
is detectable in principle.
Note that the fundamental mass parameter ${\tilde m} = \sqrt{A^2+1} m$
is of order $(100a)^{-1}$, which is rather close to the reduced Planck
scale. It might be nearly the largest mass scale
in the effective supergravity
theory, where the massive degrees of freedom as `heavy' as the reduced
Planck scale have already been integrated out.

Phenomenologically, large kinetic terms seem suitable
for not only inflaton fields but also others
such as Polonyi
\cite{Yana}
and axion fields.
We are led to suspect a possibility
that those fields reside in a single sector and the choice
of a large factor for kinetic terms is
due to inflationary reasons.

\section*{Acknowledgements}

This work is supported by the Grant-in-Aid for the 21st Century COE
"Center for Diversity and Universality in Physics"
from the Ministry of Education, Culture, Sports, Science and
Technology (MEXT) of Japan.

\end{document}